\documentstyle[12pt,aaspp]{article}
\cpright{AAS}{1994}
\lefthead{Balbes, Boyd, Steigman, \& Thomas}
\righthead{Photoerosion of Primordial Elements}
\slugcomment{OSU-TA-7/95}
\begin{document}
\title{\vspace{2.8cm}Post Big Bang Processing of the Primordial Elements}

\author{M. J. Balbes, R. N. Boyd\altaffilmark{1}, G.
Steigman\altaffilmark{1} and D. Thomas}

\affil{Department of Physics, The Ohio State University, Columbus, OH 43210}
\altaffiltext{1}{Also Department of Astronomy, Ohio State University,
			Columbus, OH 43210}
\vspace{3cm}
\begin{abstract}

We explore the Gnedin-Ostriker suggestion that a post-Big-Bang
photodissociation process may modify the primordial abundances of the light
elements. We consider several specific models and discuss the general
features that are necessary (but not necessarily sufficient) to make the
model work.  We find that with any significant processing, the final D
and $^3$He abundances, which are independent of their initial standard big
bang nucleosynthesis (SBBN) values,
rise quickly to a level several
orders of magnitude above the observationally inferred primordial values.
Solutions for specific models show that the only initial abundances
that can be photoprocessed into agreement with observations are those that
undergo virtually no processing and are already in agreement with observation.
Thus it is unlikely that this model can work for any non-trivial case
unless an artificial density and/or photon distribution is invoked.

\end{abstract}
\newpage
\section{Introduction}

The standard model of big bang nucleosynthesis (SBBN)(\cite{Pee66};
\cite{Wag67}; \cite{Sch77}; \cite{Boe85}; \cite{Wal91}) makes well defined
predictions of the primordial abundances of the light elements. These
predictions agree reasonably well with the commonly accepted values of the
primordial element abundances as inferred from observation (\cite{Wal91};
\cite{Bal93}; \cite{Smi93}) at a baryon-to-photon ratio of about $\eta \sim
3 \times 10^{-10}$. However, this baryon density corresponds to only a few
percent of the closure density of the universe. This result has motivated a
number of attempts to reconcile the observed big bang abundances with a
model in which $\eta$ is much larger.

In this context, it was suggested by \cite{Gne92}(GO) that a population of
black holes may have formed from an early generation of massive stars. The
collapse of the stars to black holes does not significantly contaminate the
primordial abundances. The black holes then accrete matter and produce a
photon bath which further processes the remaining primordial material. GO chose
initial SBBN abundances at two values of the baryon-to-photon ratio which
are higher than that which agrees with any of the abundance determinations.
Thus, they start with an excess of $^4$He and $^7$Li and
a deficit of D and $^3$He. The $^4$He and $^7$Li are then
processed into the lighter elements as a function of the total photon
energy input into the system. Although GO are not able to
process any significant amounts of $^4$He, they claim to be
in agreement with the observed primordial abundances at the 3$\sigma$ level.
(See also a  more recent paper (\cite{Gne95})).

In this paper we investigate in detail general models of nucleosynthesis
induced by post-Big-Bang photoprocessing and determine the conditions under
which the observed light element abundances can be reproduced. As we will
show, we are unable to find a non-trivial solution where the initial
abundances do not already agree with observation.  The reason for this is
straightforward.  The SBBN abundance of $^4$He is larger, by some 4 or more
orders of magnitude, that those of D and/ or $^3$He (especially at high
values of $\eta$).  It is therefore not possible to destroy enough $^4$He
to achieve consistency with observations without, at the same time, grossly
overproducing pregalactic D and/or $^3$He.  Further, at high $\eta$, $^7$Li
is significantly overproduced and it is difficult to destroy enough $^7$Li
without destroying too much $^4$He (and, consequently, overproducing D and
$^3$He).  Our detailed calculations will provide support for this simple
overview.

\newpage
\section{The Photon Spectrum and Energy-Weighted Cross Sections}
\label{section:spectrum}
In the present scenario, the primordial elements are assumed to be bathed
in a photon field for a given exposure time and photon intensity. In order
to calculate the nucleosynthesis, it is necessary to
know the energy spectrum of the radiation emitted by the black holes. Since
such spectra are not observed directly, we look to other accreting
objects, such as quasars, which may have similar spectra.
Measurements have been made of NGC 4151 (\cite{Per81}; \cite{Bai84}) and
other quasars (\cite{von87}) in energy ranges from 2 keV (\cite{Bai84}) up
to 100 MeV (\cite{Tro77}; \cite{Fic78}).

The photon number spectrum can be parameterized by two power laws, one
describing the data below 2--3 MeV and with the form $E^{-1.6}$
(\cite{Bai84}), the other describing the data above 2--3 MeV with the form
$E^{-2.7}$ (\cite{Rot83}). We are only concerned with energies above 2.2
MeV, the binding energy of the deuteron, and thus only use the higher
energy power-law. It should be noted that \cite{Gne92} give a more
complicated expression for the $\gamma$-ray spectrum in order to describe
the full energy range. As we shall see, the choice of $\gamma$-ray spectrum
does not change our conclusions.

Cross sections for ($\gamma$, X) reactions, where X = n, p, d, t, $^3$He,
or $\alpha$ have been measured over most of the energy range of interest.
In order to describe the photoerosion process, we first calculate the
energy-weighted cross sections (\cite{Boy89})
\begin{equation}
\left<\sigma\right> = \frac{\int_{2.2 MeV}^{\infty} \sigma(E) E^{-2.7} dE}
                          {\int_{2.2 MeV}^\infty E^{-2.7} dE}.
\end{equation}
Because of the power-law weighting factor, the energy-weighted cross
section is dominated by contributions from the excitation function near
threshold.  Sparse data exist for photodissociation reactions on the
lithium isotopes near threshold, contributing up to a 50\% systematic
uncertainty in the energy-weighted cross section. Cross sections for
photodissociation reactions on $^4$He, $^3$He, and D are well determined
from threshold up to several tens of MeV. The errors in these measurements
are sufficiently small ($<$ 20\%) that our conclusions will not be affected
by improved measurements of these cross sections. Table
\ref{Table:CrossSections} contains a summary of the calculated
energy-weighted cross sections.

\newpage
\section{Photoprocessing of the Elements}
\label{section:photoprocessing}
We can describe the evolution of the element abundances locally by
\begin{equation}
\label{Eqn:dNdt}
\frac{1}{\phi}
\frac{d\left(N_i\right)}{dt} =
	\sum_j\left(N_j\right)\left<\sigma_{j + \gamma \rightarrow i}\right>
	    - \left(N_i\right)\left<\sigma_{i + \gamma \rightarrow j}\right>
\end{equation}
where $N_i = N_i(\vec{r},t)$ is the number density (at location $\vec{r}$ and
time t) of nuclide $i$ and
$\phi = \phi(\vec{r})$ is the time-independent
photon flux. We can further generalize Eqn. \ref{Eqn:dNdt} by writing
the number density as $N_i(\vec{r},t) =
f_i\left(\phi(\vec{r})t\right)
\rho(\vec{r})$, where $\rho(\vec{r})$ is the local matter density.
For a time-independent $\rho(\vec{r})$, Eqn. \ref{Eqn:dNdt} can be rewritten
with the $f_i$ replacing the $N_i$ and yielding solutions which are the same to
within a dimensional constant.

In general, Eqn. \ref{Eqn:dNdt} might also include
$\beta$-decay terms which act as both production and destruction
mechanisms. In our case the
only unstable nuclides of
importance are $^3$H and neutrons. Reaction cross sections (and thus the
energy-weighted cross sections) for photoproduction and photodissociation
of $^3$H are very similar to those for $^3$He (see Table
\ref{Table:CrossSections}). For neutrons the situation is even simpler
since all are assumed to decay to protons without undergoing nuclear
reactions. Therefore we treat $^3$H as
$^3$He and neutrons as protons in our calculations.

The simplest model is to adopt a uniform isotropic
$\gamma$-ray
distribution (i.e. $\phi$ is constant) and a uniform matter density
distribution in Eqn. \ref{Eqn:dNdt}.  Such a distribution could be
produced by a universe populated with a high density of black holes which
tend to smear out the $1/r^2$ dependence of the photon flux from an individual
source. Eqn.
\ref{Eqn:dNdt} then describes a set of coupled differential equations which
can be evolved forward in time as a function of the "exposure"($\phi t$) for
each
element. Furthermore, the solutions obtained from this simple model can be
modified to include effects from non-isotropic photon fluxes or non-uniform
density distributions. These effects are folded into the solutions to Eqn.
\ref{Eqn:dNdt} by a weighting procedure of the form

\begin{equation}
\label{Eqn:weighted_average}
\left<\frac{N_i}{N_H}\right> =
\frac{\int_0^\infty w(x)N_i(x) dx}
     {\int_0^\infty w(x)N_H(x) dx}
\end{equation}
where $x=\phi(\vec{r})t$ is the position-dependent exposure.
Note that with $w(x) = \delta(x-\phi(\vec{r}) t)$ we recover the
original local solutions.

We examine the effects of several functional forms for $w(x)$. In the first
case we assume that there are two separately processed regions. Then $w(x)$
is simply a sum of two $\delta$ functions. In the second case, we have
assumed a uniform population of black holes surrounded by a uniform density
of matter. Processing of the matter is dominated by the nearest black hole.
Thus the final element abundances are given by a spatial averaging of the
processed material. In the third case, we consider not only a distribution
of black holes but a density distribution which varies as $1/r^2$ so as to
produce a flat rotation curve. From these examples, we reach some general
conclusions about the necessary (and artificial) form $w(x)$ must have in
order to obtain a viable model.

The criterion for these models to be viable is that the calculated
abundances are within the limits determined by the observed light element
abundances (see Table \ref{Table:ObsLimits}). We have chosen realistic but
generous (2$\sigma$ or greater) limits so that if a model does fail, the
failure is due to the model not to the choice of limits. The standard model
of big bang nucleosynthesis is in agreement with our limits for $2.5 \times
10^{-10} < \eta < 4.6 \times 10^{-10}$.

In each of the scenarios mentioned above, we have presented the solutions
in two different ways. First, we use initial abundances from standard big
bang nucleosynthesis which allows us to parameterize our four initial
abundances in terms of the baryon-to-photon ratio $\eta$. We can then
follow the evolution of the abundances as a function of the exposure, $\phi
t$. Second, since it can be argued that an assumption of the existence of
primordial black holes negates the SBBN model and necessitates the use of
the inhomogeneous model, we invert the solutions to
Eqn. \ref{Eqn:dNdt} for each of our scenarios. Then using the observed
abundances, we can trace the evolution backwards in time to solve for the
primordial values as a function of the exposure. Thus we avoid making
any assumptions about the isotopic content of the pre-photoprocessed
universe.

\section{Results}
\label{section:results}
\subsection{Isotropic photon flux}

We have calculated the abundances of the light elements relative to
hydrogen for isotropic exposures $0 \leq \phi t \leq 10^3$
photons/$10^{-27}$ cm$^2$ starting with initial abundances determined from
the SBBN model with $10^{-10}<\eta<10^{-8}$. Fig. \ref{Fig:abun_vs_phit}
shows the evolution of the abundances as a function of the exposure, $\phi
t$. Horizontal lines represent the observationally inferred abundances. In
Fig. \ref{Fig:abun_map} we map out the regions of $\phi t$ versus $\eta$
space where the individual abundances agree with observation. At no values
of $\eta$ and $\phi t$ can the processed abundances of $^4$He, $^3$He, and
D be reconciled with the observed limits except in the region where the
unprocessed abundances already essentially agree with observation. In fact,
this scenario can be ruled out by just looking at the $^4$He and $^3$He
abundances. Since most $^4$He is being converted directly into $^3$He and
the $^4$He abundance is so much greater than $^3$He, any small change in
the $^4$He abundance is accompanied by a very large change in the $^3$He
abundance. This is clearly seen in the figures to be independent of the
initial $^3$He abundance (and therefore also independent of $\eta$).
Deuterium is constrained to rise more slowly than $^3$He because its
production is predominantly from $^3$He through the $^3$He($\gamma$,p)D
reaction. Therefore, deuterium production cannot proceed until a
significant quantity of $^3$He has been created.

At large exposures, when most of the $^4$He has been destroyed, the $^3$He
abundance begins to decrease, eventually reaching agreement with
observation. However, this large $\phi t$ scenario is ruled out because the
$^4$He and D abundances are well outside the observational bounds.

We have inverted the solutions to Eqn. \ref{Eqn:dNdt} in order to calculate
the possible combinations of initial abundances for which this model will
work. We are therefore no longer constrained by SBBN.
Solutions for the initial abundances, which when processed for a
given $\phi t$ yield light element abundances in agreement with observations,
are found only for $\phi t < 9.1 \times 10^{-3}$
photons/$10^{-27}$ cm$^2$. Therefore, no significant processing
occurs. Indeed, the {\em initial} $^4$He and $^3$He abundances are
constrained to be within the observational limits. The initial D
abundance can be less than the observed lower limit by only $\sim$ 5\%
and cannot be more than the observed upper limit. The initial
$^7$Li abundance can be no more than 0.3\%  above the upper limit.

\subsection{Two-zone model}

We have investigated the possibility of reconciling the helium and
deuterium isotope abundances by assuming a two-zone model, i.\ e.\ $w(x)$
is a sum of two $\delta$ functions. This provides a
simplified description of a mixture of material processed through different
exposures. In this case, space is divided into two regions, one of
which undergoes very little or no processing and the other undergoes a
great deal of processing. In the former case, deuterium and $^3$He agree
with observation and $^4$He is overabundant. It is also necessary that
$\eta < 10 \times 10^{-10}$ so that deuterium is not underproduced (see
Fig. \ref{Fig:abun_vs_phit}). The latter zone undergoes significantly more
processing such that most of the $^4$He is destroyed. Also, the $^3$He and
deuterium which have been created from the processed $^4$He have been
destroyed to such an extent that they once again are close to the
observational limits. Then by adjusting the relative volumes of the two
zones (such that the larger volume is only slightly processed and the
smaller volume is highly processed), one can force agreement between the
processed $^4$He and the observational limits. In the larger
mostly-unprocessed zone the $^3$He abundance constrains $\phi t <
10^{-2}$. The $^4$He and $^7$Li are therefore essentially unprocessed.
However, since $^7$Li is overproduced by a greater factor than $^4$He, and
the zone-mixing reduces the $^7$Li by the same fraction as $^4$He, the
$^7$Li is still overproduced in this model. The only way to reconcile the
model with observation is to start with primordial abundances for $^7$Li,
$^3$He, and deuterium which agree with the observational limits already.
Even then, the primordial $^4$He cannot be more than a few percent
overabundant due to the constraints placed on $\eta$ by $^7$Li.

\subsection{Uniform black hole distribution}
Although the two-zone model shows clearly why agreement cannot be attained
 between the primordial abundances and observation,
it is too simple a model from which to infer general conclusions. In
order to investigate a more physical picture, we have studied a model
that has a uniform distribution of black holes that are assumed to be weak
enough and/or far enough apart that processing of the primordial material
is dominated by the closest black hole. In order to account for the spatial
variation of the photon flux, we modify Eqn. \ref{Eqn:dNdt} by setting
$\phi(r) = L_{\gamma}/4\pi r^2$, where $L_{\gamma}$ is the total
$\gamma$-ray luminosity from one accreting black hole. We can then write the
spatially-averaged abundance relative to hydrogen as
\begin{equation}
\label{Eqn:radial_average}
\left<\frac{N_i}{N_H}\right> =
\frac{\int_{R_{min}}^{R_{max}} N_i\left(\frac{L_{\gamma} t}
		{4\pi r^2}\right) r^2 dr}
     {\int_{R_{min}}^{R_{max}} N_H\left(\frac{L_{\gamma} t}
		{4\pi r^2}\right) r^2 dr}
=
\frac{\int_{R_{min}}^{R_{max}} f_i\left(\frac{L_{\gamma} t}
		{4\pi r^2}\right) \rho(r) r^2 dr}
     {\int_{R_{min}}^{R_{max}} f_H\left(\frac{L_{\gamma} t}
		{4\pi r^2}\right) \rho(r) r^2 dr}.
\end{equation}

We examine two cases in detail. In the first case, we assume that $\rho(r)$
is constant, i.e. the matter is distributed homogeneously. Then Eqn.
\ref{Eqn:radial_average} can be written more conveniently in terms of one
parameter $x = \frac{L_{\gamma} t}{4\pi r^2}$ to yield
\begin{equation}
\label{Eqn:radial_multsqrtx}
\left<\frac{N_i}{N_H}\right> =
\frac{\int_{x_{min}}^{x_{max}} f_i\left(x\right) x^{-5/2} dx}
     {\int_{x_{min}}^{x_{max}} f_H\left(x\right) x^{-5/2} dx}.
\end{equation}

Alternatively, if we assume that the density of matter falls off with
distance from the central source as $\rho(r) =
\rho_0(\frac{R}{r})^2$ (which will yield the flat rotation curves seen in
galaxies) then Eqn. \ref{Eqn:radial_average} becomes
\begin{equation}
\label{Eqn:radial_divsqrtx}
\left<\frac{N_i}{N_H}\right> =
\frac{\int_{x_{min}}^{x_{max}} f_i\left(x\right) x^{-3/2} dx}
     {\int_{x_{min}}^{x_{max}} f_H\left(x\right) x^{-3/2} dx}.
\end{equation}

Figures \ref{Fig:uniform_density} and \ref{Fig:flat_rotation} show the
dependence of the abundances on $x_{min}$ (with $x_{max}=\infty$ so that
$R_{min}=0$) for three
values of $\eta$. Figures \ref{Fig:uniform_density_map} and
\ref{Fig:flat_rotation_map} illustrate the regions of $\eta-x_{min}$ space
where the spatially-averaged abundances agree with observation. It is
clearly seen that there is no region of parameter space where the models
agree with the observed abundances of $^4$He, $^3$He, and D simultaneously
except where the standard model (for low baryon density) already works. We
have also searched for solutions for $x_{min}<x_{max}<\infty$, where
processed material with exposure greater than $x_{max}$ is assumed to be
swallowed by the black hole and thus does not contribute to the spatial
average. No non-trivial solutions were found.

As with the isotropic photon flux scenario, we have inverted the
solutions to Eqns. \ref{Eqn:radial_multsqrtx} and \ref{Eqn:radial_divsqrtx}
in order to calculate the possible combinations of initial abundances for which
these models will succeed. Solutions for the initial abundances, which
when processed yield light element abundances in
agreement with observations, are found only for $\phi t < 3 \times 10^{-3}$
and $\phi t < 8 \times 10^{-6}$ photons/$10^{-27}$ cm$^2$ for the uniform
density and the flat rotation curve models, respectively.
Again, no significant processing occurs and the {\em initial}
$^4$He and $^3$He abundances are constrained to be within the
observed limits. The initial D abundance may be less than the
observed lower limit by only $\sim$ 5\% and it cannot be higher than the
observed upper limit. The initial $^7$Li abundance can be no more than
0.3\% above the upper limit.

It might be thought that the detailed shape of the photon spectrum
will affect the present results. However, changing the power law index or
using the parameterization of GO causes only small changes in the
energy-weighted cross sections. In the GO case, the
energy-weighted cross sections are increased by no more than a factor of 4
even when using the most extreme distortion of the spectrum. Furthermore,
changes in the spectra change all the energy-weighted cross sections in the
same direction, so the effect on resulting abundances is less than
that on the cross sections. Since our conclusions are based on qualitative
conflicts between observations and the model predictions, it is not likely
that uncertainties in the photon spectrum will affect our conclusions.

\subsection{A general solution?}

So far we have provided examples of models that don't work. Is it possible
to construct a non-trivial model that will work? That is, can we
construct a weighting function $w(x)$ (see Eqn. \ref{Eqn:weighted_average})
 that will reconcile the solutions to Eqn. \ref{Eqn:dNdt}
(shown in Fig. \ref{Fig:abun_vs_phit}) with the observed
limits? In constructing such a model, two important features must be kept
in mind. First, the weighting function must preferentially mix regions such
that a larger fraction of $^7$Li is destroyed than of $^4$He. Since the
destruction cross section for $^7$Li is higher than for $^4$He, this may be
possible. Second, the weighting function must avoid appreciable
contributions from the region $10^{-2} < \phi t < 5 \times 10^2$ in order
to avoid overproducing $^3$He and D. As a result, such a function is
forced to be of the form shown in Fig. \ref{Fig:weight}. This bimodal
distribution is decidedly ad hoc! But, even with a weighting
function of this type, it is difficult (and perhaps impossible) to
reproduce the observed primordial abundances of D, $^3$He, $^4$He, and
$^7$Li.

\section{Conclusions}
\label{section:conclusions}

We have investigated several scenarios in which the primordial abundances
of the light elements undergo photoprocessing due to a population of
$\gamma$-emitting objects such as accreting black holes (GO). In all cases,
the primordial $^4$He abundance is constrained to agree with observation since
even a small amount of processing of $^4$He increases the $^3$He and D
abundances by several orders of magnitude. At large exposures, when
destruction of $^3$He and D has occurred, their abundances relative to
hydrogen become independent of initial values and cannot simultaneously be
reconciled with observation. A general solution which would correct for
this effect must preferentially weight the processed regions in such a way
as to be pathological.

In short, we find no non-trivial solution to any of the several
photodissociation models considered for either SBBN or inhomogeneous BBN
abundances corresponding to
large values of $\eta$. Post BBN photoprocessing of the light elements
cannot weaken the upper bound on the universal density of baryons.

\acknowledgments
This work was supported in part by National Science Foundation grant
PHY92-21669 and Department of Energy grant DE-AC02-76ER01545.

\begin{figure}
\caption{Abundances relative to hydrogen of the light elements as functions of
exposure,
$\phi t$. The initial
abundances are calculated from SBBN with $\eta_{10} = 3$
(solid curve), $\eta_{10} = 10$ (dashed curve), and $\eta_{10} = 100$
(dashed-dotted curve). Horizontal lines indicate observational limits on the
primordial abundances.}
\label{Fig:abun_vs_phit}
\end{figure}

\begin{figure}
\caption{Parameter space for the isotropic photon flux model. Hatched areas
indicate agreement between the calculated and observed elemental abundance.
Vertical hatching is D, hatching with a positive slope is $^3$He, and
hatching with a negative slope is $^4$He. The region where $^7$Li agrees
with observation is not shown, however it also overlaps the region of
agreement between the other three light elements.}
\label{Fig:abun_map}
\end{figure}

\begin{figure}
\caption{Abundances relative to hydrogen of the light elements when
spatially averaged over a
uniform density distribution and a photon flux which decreases as 1/r$^2$.
The abscissa is $x_{min}$, where $x = Lt/4\pi R^2$.
The initial abundances are calculated from SBBN with $\eta_{10} = 3$ (solid
curve), $\eta_{10} = 10$ (dashed curve), and $\eta_{10} = 100$
(dashed-dotted curve). Horizontal lines indicate observational limits on
the primordial abundances.}
\label{Fig:uniform_density}
\end{figure}

\begin{figure}
\caption{Abundances relative to hydrogen of the light elements when
spatially averaged over a density distribution and a photon flux both of
which decrease as 1/r$^2$. The abscissa is $x_{min}$, where $x = Lt/4\pi
R^2$. The initial abundances are calculated from SBBN with $\eta_{10} = 3$
(solid curve), $\eta_{10} = 10$ (dashed curve), and $\eta_{10} = 100$
(dashed-dotted curve). Horizontal lines indicate observational limits on
the primordial abundances.}
\label{Fig:flat_rotation}
\end{figure}

\begin{figure}
\caption{Parameter space for the spatially-averaged model with a uniform
density distribution. Hatched regions indicate where the calculated
abundances of D, $^3$He, and $^4$He agree with observation. Vertical
hatching is D, hatching with a positive slope is $^3$He, and hatching with
a negative slope is $^4$He. The only agreement for all three abundances
occurs in the region of parameter space where the standard (low baryon
density) model already works. The region where $^7$Li agrees with
observation is not shown, however it also overlaps the region of agreement
between the other three light elements.}
\label{Fig:uniform_density_map}
\end{figure}

\begin{figure}
\caption{Parameter space for the spatially-averaged model with a matter
density which varies as $1/r^2$. Hatched regions indicate where
the calculated abundances of D, $^3$He, and $^4$He agree with observation.
Vertical hatching is D, hatching with a positive slope is $^3$He, and
hatching with a negative slope is $^4$He. The only agreement for all three
abundances occurs in the region of parameter space where the standard (low
baryon density) model already works. The region where $^7$Li agrees with
observation is not shown, however it also overlaps the region of agreement
between the other three light elements.}
\label{Fig:flat_rotation_map}
\end{figure}

\begin{figure}
\caption{Possible weighting function $w(x)$ (thick line)
superimposed over the
abundance vs.\ exposure curve for $^3$He.}
\label{Fig:weight}
\end{figure}

\begin{table}
\caption{Observational limits on primordial abundances.}
\begin{tabular}{lll}
		&   Abundance Limit	& Reference \\ \tableline
$^7$Li/H 	&	$< 7 \times 10^{-10}$	&	1, 2, 3\\
Y$_p$		&	0.215 -- 0.244		&	1, 2, 4, 5\\
$^3$He/H	&	$< 2 \times 10^{-5}$	&	1\\
D/H		&	1.2 -- 10 $\times 10^{-5}$ &	1, 2\\
\end{tabular}
\tablerefs{1) \cite{Wal91}, 2) \cite{Smi93}, 3) \cite{Pin92},
4) \cite{Bal93} 5) \cite{Oli95}}
\label{Table:ObsLimits}
\end{table}

\begin{table}
\caption{Energy-weighted cross sections}
\tablerefs{1) \cite{Die88}, 2) \cite{Jun79}, 3) \cite{Gre62},
4) \cite{Gri61}, 5) \cite{Bur87}, 6) \cite{Sko79}, 7) \cite{Rob81},
8) \cite{Fel90}, 9) \cite{War81}, 10) \cite{Ber80}, 11) \cite{Ber88}
12) \cite{Bar87}, 13) \cite{Bal79}, 14) \cite{Fau81}, 15) \cite{Tic73}
16) \cite{Seg64}}
\begin{tabular}{lcl}
Reaction   &   $<\sigma>$ (mb)	& Reference \\ \tableline
$^7$Li($\gamma$, n)$^6$Li	&	0.0486	&	1\\
$^7$Li($\gamma$, p)$^6$He	&	0.0480	&	2, 3\\
$^7$Li($\gamma$, d)$^5$He	&	0.0118	&	2\\
$^7$Li($\gamma$, t)$^4$He	&	0.1760	&	2, 4, 5, 6\\
\\
$^6$Li($\gamma$, n)$^5$Li	&	0.1748	&	1\\
$^6$Li($\gamma$, p)$^5$He	&	0.4334	&	2\\
$^6$Li($\gamma$, d)$^4$He	&	0.0047	&	7\\
$^6$Li($\gamma$, t)$^3$He	&	0.0147	&	2\\
\\
$^4$He($\gamma$, n)$^3$He	&	0.0153	&	8, 9, 10\\
$^4$He($\gamma$, p)$^3$H	&	0.0167	&	8, 11\\
$^4$He($\gamma$, d)D	&	$7.7 \times 10^{-5}$	&	12\\
$^4$He($\gamma$, pn)D	&	0.0013	&	13\\
\\
$^3$He($\gamma$, n)pp		&	0.0547	&	14\\
$^3$He($\gamma$, p)D		&	0.0913	&	15\\
\\
$^3$H($\gamma$, n)D		&	0.0547	&	14\\
$^3$H($\gamma$, p)nn		&	0.0788	&	14\\
\\
D($\gamma$, p)n		&	1.7473	&	16

\end{tabular}
\label{Table:CrossSections}
\end{table}


\newpage
\begin{thebibliography}{}

\bibitem[Baity et al.\ 1984]{Bai84}
\reference Baity, W. A., Mushotzky, R. F., Worrall, D. M., Rothschild, R. E.,
	Tennant, A. F., Primini, \& F. A. 1984, \apj, 279, 555

\bibitem[Balbes, Boyd, \& Mathews 1993]{Bal93}
\reference Balbes, M. J., Boyd, R. N., \& Mathews, G. J. 1993, \apj, 418, 229

\bibitem[Balestra et al.\ 1979]{Bal79}
\reference Balestra, F., Busso, L., Garfagnini, R., Piragino, G.,
	\& Zanini, A. 1979, Nouvo Cimento, 49, 575

\bibitem[Barnes et al.\ 1987]{Bar87}
\reference Barnes, C. A., Chang, K. H., Donoghue, T. R., Rolfs, C., \&
	Kammeraad, J. 1987, Phys. Lett. B, 197, 315

\bibitem[Berman 1980]{Ber80}
\reference Berman, B. L., Faul, D. D., Meyer, P., \& Olson, D. L. 1980,
	\prc, 22, 2273

\bibitem[Bernabei 1988]{Ber88}
\reference Bernabei, R., Chisholm, A., d'Angelo, S., de Pascale, M., P.,
             Picozza, P., Shaerf, C., Belli, P., Casano, L., Incicchitti, A.,
             Prosperi, D., \& Girolami, B. 1988, \prc, 38, 1990

\bibitem[Boesgaard \& Steigman 1985]{Boe85}
\reference Boesgaard, A. M. \& Steigman, G. 1985, ARA\&A, 23, 319

\bibitem[Boyd, Ferland \& Schramm 1989]{Boy89}
\reference Boyd, R. N., Schramm, D. N., \& Ferland, G. J. 1989, \apj, 336, L1

\bibitem[Burzynski et al.\ 1987]{Bur87}
\reference Burzynski, S., Czerski, K., Marcinkowski, A., \& Zupranski, P. 1987,
	Nucl. Phys., A473, 179

\bibitem[Dietrich \& Berman 1988]{Die88}
\reference Dietrich, S. S., \& Berman, B. L. 1988, Atomic Data and Nuclear
	Data Tables, 38, 199

\bibitem[Faul et al.\ 1981]{Fau81}
\reference Faul, D. D., Berman, B. L., Meyer, P., \& Olson, D. L. 1981,
	\prc, 24, 849

\bibitem[Feldman et al.\ 1990]{Fel90}
\reference Feldman, G., Balbes, M. J., Kramer, L. H., Williams, J. Z.,
	Weller, H. R., \& Tilley, D. R. 1990, \prc, 42, R1167

\bibitem[Fichtel, Simpson, \& Thompson 1978]{Fic78}
\reference Fichtel, C. E., Simpson, G. A. \& Thompson, D. J. 1978,
		\apj, 222, 833

\bibitem[Gnedin \& Ostriker 1992]{Gne92}
\reference Gnedin, N. Y., \& Ostriker, J. P. 1992, \apj, 400, 1

\bibitem[Gnedin, Ostriker \& Rees 1995]{Gne95}
\reference Gnedin, N. Y., Ostriker, J. P., \& Rees, M. J. 1995, \apj, 438, 40

\bibitem[Gregory, Sherwood, \& Titterton 1962]{Gre62}
\reference Gregory, A. G., Sherwood, T. R., \& Titterton, E. W. 1962,
	Nucl. Phys., 32, 543

\bibitem[Griffiths et al.\ 1961]{Gri61}
\reference Griffiths, G. M., Morrow, R. A., Riley, P. J., \& Warren, J. B.
1961,
	Can. J. Phys., 39, 1397

\bibitem[Junhgans et al.\ 1979]{Jun79}
\reference Junghans, G., Bangert, K., Berg, U. E. P., Stock, R., \& Wienhard,
		K. 1979, Z. Phys., A291, 353

\bibitem[Olive \& Steigman 1995]{Oli95}
\reference Olive, K. A. \& Steigman, G. 1995, \apjs, 97, 49

\bibitem[Peebles 1966]{Pee66}
\reference Peebles, P.J.E. (1966) Phys. Rev. Lett. 16, 410
\bibitem[Perotti et al. 1981]{Per81}
\reference Perotti, F., Della Ventura, A., Villa, G., Di~Cocco, G., Bassani,
	L., Butler, R. C., Carter, J. N., \& Dean, A. J. 1981, \apj, 247, L63

\bibitem[Pinsonneault, Deliyannis, \& Demarque 1992]{Pin92}
\reference Pinsonneault, M. H., Deliyannis, C. P. \& Demarque, P. 1992,
\apjs, 78,179

\bibitem[Robertson et al.\ 1981]{Rob81}
\reference Robertson, R. G. H., Dyer, P., Warner, R. A., Melin, R. C.,
	Bowles, T. J., McDonald, A. B., Ball, G. C., Davies, W. G.,
	\& Earle, E. D., 1981, \prl, 47, 1867

\bibitem[Rothschild et al.\ 1983]{Rot83}
\reference Rothschild, R. E., Mushotzky, R. F., Baity, W. A., Gruber, D. E.,
	 Matteson, J. L.,  \& Peterson, L. E. 1983, \apj, 269, 423

\bibitem[Schramm \& Wagoner 1977]{Sch77}
\reference Schramm, D. N., \& Wagoner, R. V. 1977, Ann.\ Rev.\ Nucl.\ Part.
	\ Sci.\, 27, 37

\bibitem[Segr\`{e} 1964]{Seg64}
\reference Segr\`{e}, E. 1964, Nuclei and Particles (New York, NY:
	W. A. Benjamin, Inc.)

\bibitem[Skopik et al.\ 1979]{Sko79}
\reference Skopik, D. M., Asai, J., Tomusiak, E. L., \& Murphy II, J. J. 1979,
	\prc, 20, 2025

\bibitem[Smith, Kawano, \& Malaney 1993]{Smi93}
\reference Smith, M. S., Kawano, L. H., \& Malaney, R. A. 1993, \apjs, 85, 219

\bibitem[Songaila et al. 1994]{Son94}
\reference Songaila, A., Cowie, L. L., Hogan, C. J., \& Rugers, M. 1994,
		Nature, 368, 599

\bibitem[Ticcioni et al.\ 1973]{Tic73}
\reference Ticcioni, G., Gardiner, S. N., Matthews, J. L., \& Owens, R. O.
	1973, Phys. Lett., 46B, 369

\bibitem[Trombka et al.\ 1977]{Tro77}
\reference Trombka, J. I., Dyer, C. S. Evans, L. G., Bielefeld, M. J., Seltzer,
	 S. M.,  \& Metzger, A. E. 1977, \apj, 212, 925

\bibitem[von Ballmoos, Diehl, and Schonfelder 1987]{von87}
\reference von Ballmoos, P., Diehl, R., \& Schonfelder, V. 1987, \apj, 312, 134

\bibitem[Wagoner, Fowler, \& Hoyle 1967]{Wag67}
\reference Wagoner, R. V., Fowler, W. A., \& Hoyle, F. 1967, \apj, 148,3

\bibitem[Walker et al.\ 1991]{Wal91}
\reference Walker, T. P., Steigman, G., Schramm, D. N., Olive, K. A.,
	\& Kang, H. 1991, \apj, 376, 51

\bibitem[Ward et al.\ 1981]{War81}
\reference Ward, L., Tilley, D. R., Skopik, D. M., Roberson, N. R., \&
	Weller, H. R. 1981, \prc, 24, 317

\end{thebibliography}
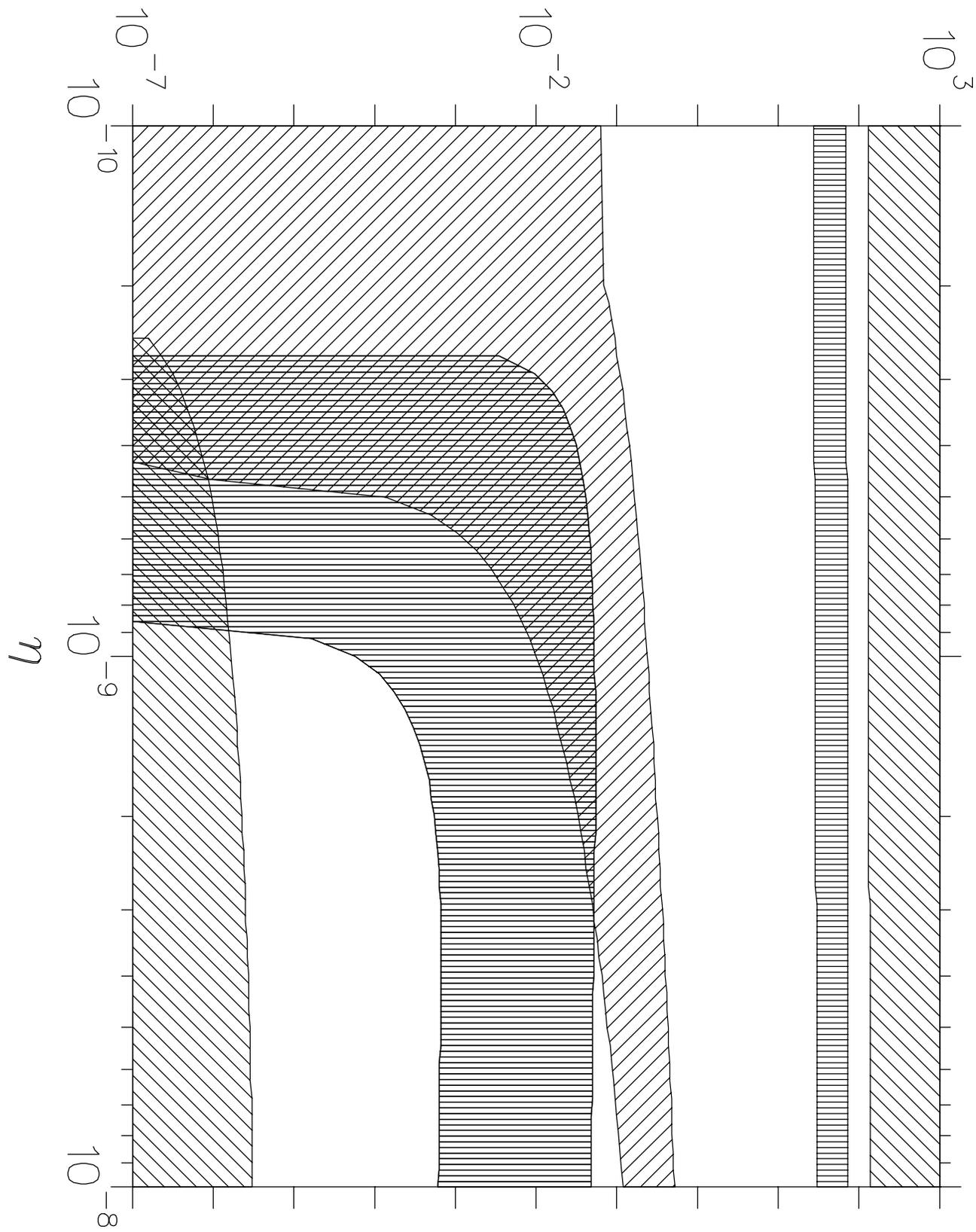
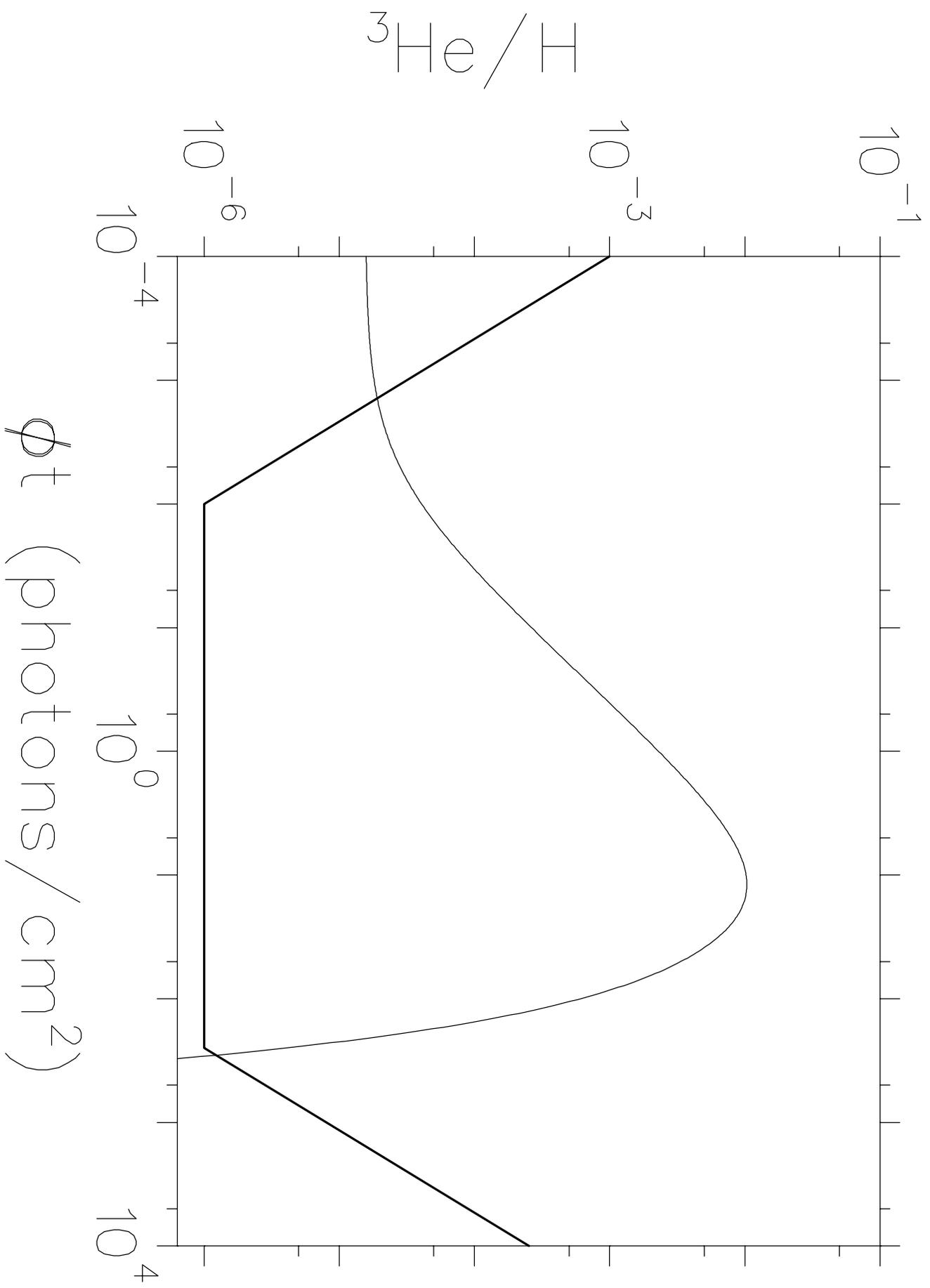
\end{document}